\documentclass{article}

\usepackage{graphicx}

 \usepackage[textwidth=3.4cm,textsize=10pt]{todonotes}

\newtheorem{definition}{Definition}

\makeatletter \let\c@table\c@figure \makeatother

\title{Delegated Causality of Complex Systems} 
\author{Raimundas Vidunas\\
\em Osaka University}

\renewcommand{\baselinestretch}{1.005} 
\let\OLDthebibliography\thebibliography
\renewcommand\thebibliography[1]{
  \OLDthebibliography{#1}
  \setlength{\parskip}{0pt}
  \setlength{\itemsep}{0pt plus 0.3ex}
}

\newcommand{\refpart}[1]{{\it (#1)}}  

\begin{document}

\date{}
\maketitle

\begin{abstract}
A notion of {\em delegated causality} is introduced here.
This subtle kind of causality is dual to interventional causality. 
Delegated causality elucidates 
the causal role of dynamical systems at the ``edge of chaos",
explicates evident cases of downward causation,
and relates emergent phenomena to G\"odel's incompleteness theorem.
Apparently rich implications are noticed 
in biology and Chinese philosophy. 
\end{abstract}
 

\section{Introduction}

Living organisms, ecosystems, human minds,  
societies, economic markets are widely recognized as extraordinary complex systems. 
They are impressively 
organized and possess properties that are hardly reducible to qualities of physical matter.
Thereby 
they seem to contradict the reductionistic paradigm of fundamental causation 
from underlying physical processes. As yet, 
satisfying explanation of emerging coherent organization is 
a comparable 
challenge for reductionist and holistic philosophies 
\cite{Capra15}, \cite{HeyCG}.
Even if the reductionist approach continues to deliver 
outstanding 
results in physics, chemistry, molecular biology, neuropsychology,
much deeper understanding of 
living \cite{Shrod44}, \cite{ShrodMore}  
and conscious \cite{Kim98}, \cite{VarelaTR}
agencies may require an uneasy 
 paradigm change 
\cite{Kuhn70}, after all.

I introduce a concept that can simplify and unify analysis 
of intricate causal relations in complex systems 
to a remarkable extent. 
This concept of {\em delegated causality} should clarify much 
about emergence of whole new phenomena \cite{Emergence06},
spontaneous order \cite{Kauffman93},  
synergy \cite{Corning05}, functionality \cite{Functions02}, 
purpose and intention \cite{Dennett87}.
If the new conception 
indeed refines established specialist perspectives,
it will be worth revisiting sporadic revivals of Emergentism 
\mbox{\cite[p.~9--26]{Emergence06}},
post-Enlightenment rationalist skepticism \mbox{\cite[p.~114]{Emergence06}}, 
classical Greek teleologies \mbox{\cite[p.~7--30]{Functions02}}. 
The most rigorous contemporary relevance of the new perspective
is to physics of emergence \cite{Mainwood06},
symmetry breaking \cite{Anderson72},  \cite{MoonLR17}, 
thermodynamics \cite{Prigogine77}, \cite{England13}, 
and to information-theoretic measure of causal influence \cite{Hoel17}, \cite{TononiSporns03}. 

A comprehensive overview of the vast, 
growing literature on complex systems, self-organization, 
emergence 
would not serve the purpose of this article to introduce
delegated causality. 
This simple but subtle, overlooked kind of causality is 
{\em provoked} (figuratively speaking)
by critical dynamical systems with rich 
behavior and moderate sensitivity to the environment.
The scope of my abstracted terminology will become clear with 
the introduction of methodology \refpart{M1}--\refpart{M3} in \S \ref{sec:delegate}
of analyzing causal interactions. 
Evident implications of {delegated causality} will be demonstrated 
by a brief account 
of evolutionary biology (in \S \ref{sec:biology})
and a reference to Chinese 
philosophy (in \S \ref{sec:yinyang}). 

This spirited article would be presentable to 
a scientific version of the TV show {\em ``The X-Factor"} \cite{xfactor}. 
My argumentation is not deep formally, as the chief 
purpose is to justify the new concept by a few evocative arguments, 
agreeable examples, and links 
to existing ideas.
I start by reassessing contemporary 
modeling of complex systems in \S \ref{sec:csystems}. 
The fresh kind of causality is introduced formally in \S \ref{sec:delegate}.
Section 4 examines physical reductionism in the new light,  
and relates emergence, downward causation 
to G\"odel's incompleteness theorem \cite{Godel31}. 
The later sections deliberate 
a few compelling 
(though not entirely comfortable) implications. 
All together, this article is gradually making a holistic argument
for a new comprehensive view  
by building up the context for the integrating Section \ref{sec:integrate}.

\section{Complex systems}
\label{sec:csystems}

Natural complex systems are studied under many 
frameworks: self-organization \cite{Ashby62}, \cite{selforganize}, 
complex adaptive systems \cite{CAS07}, \cite{Horgan95}, 
autopoiesis \cite{Autopoiesis}, dissipative structures \cite{Prigogine77}, 
self-organized criticality \cite{Bak96}, \cite{SOC15}, etc.
The models are often based on non-linear dynamical systems, their attractors,
non-equilibrium thermodynamics \cite{England13}, \cite{KKD15},
phase transitions \cite{PHSOC16}, scaling analysis \cite{West17},
cellular automata \cite{Wolfram02}, \cite{Langton90}, 
variation and selection mechanisms \cite{Heylighen00}, \cite{VSD13}, 
systems theory \cite{Bertalanffy}, \cite{Capra15},
developmental frameworks \cite{Salthe93}, \cite{Coffman},
information dynamics \cite{Lerner07}.
Phenomenology 
at the center of attention includes
spontaneous increase of order \cite{Holland95}, \cite{Kauffman93}, 
emergence of coherent global behaviors from local interactions \cite{Popkin16},
adaptation to environment perturbations. 
Applicable distinction between self-organization and emergence
is put forward in \cite{WolfHolvert05}. 

Autonomy, decentralized control, interactive closure
are often among defining features of complex systems \cite{Autopoiesis}, \cite{MorenoMossio}. 
But the autonomy assumption should not be idealized, 
especially when considering causation in actual complex phenomena.
Reflections in this article suggest that behaviors (of constituting agents or the whole system)
which amount to control sharing or transfer can be very 
far reaching.

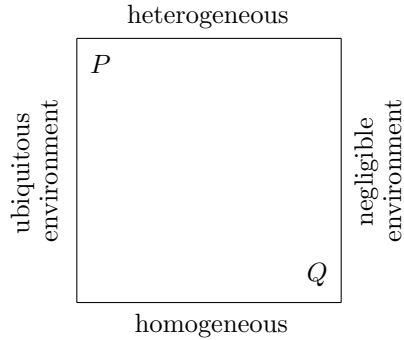
\begin{figure}
\[
\begin{picture}(120,120)(0,0)
\put(10,10){\line(1,0){100}}
\put(10,10){\line(0,1){100}}
\put(10,110){\line(1,0){100}}
\put(110,10){\line(0,1){100}}
\put(15,97){$P$}  \put(97,18){$Q$} 
\put(32,-1){homogeneous} \put(29,116){heterogeneous}
\put(-15,36){\rotatebox{90}{ubiquitous}} 
\put(115,39){\rotatebox{90}{negligible}} 
\put(-4,33){\rotatebox{90}{environment}} 
\put(126,33){\rotatebox{90}{environment}} 
\end{picture}
\]
\caption{The two-dimensional 
spectrum of interactivity with the environment and homogeneity of complex systems.}
\label{fg:2d}
\end{figure}

It will help clarity here to contrast various complex systems
on the two-dimensional spectrum that combines
interactivity with the environment and homogeneity of constituents;
see Figure \ref{fg:2d}.
Living organisms are highly non-homogeneous and experience variable 
pressure (often self-inflicted) from the environment. 
Their biological and physical organization is nested hierarchical \cite{Salthe12}:
\[ \rm
[\;organism \; 
[ \; organs \;
[ \; tissues \; 
[ \; cells \;
[ \; biomolecules \;
[ \; atoms \; [ \; \ldots \;
]\,]\,]\,]\,]\,]\,];
\]
and they live in similarly nested hierarchical environments:
\[ \rm
[\,[\,[\,[\,[\,[\,[
\; organisms \; ] 
\; flocks \; ]
\; habitats \; ]
\; ecosystems \; ]
\; biosphere \; ]
\; planet \; ]
\; \ldots \; ].
\]
General heterogeneous adaptive systems are mainly organized
in a hierarchal way as outlined by Simon \cite{Simon62}.
Intense, diverse interactivity between levels facilitates growth and adaptation.
Let $P$ denote the spectral corner representing these systems.

Let $Q$ denote the opposite spectral corner in Figure \ref{fg:2d}.
Do we find phase transitions of homogeneous matter there?
Their unfolding depends on a few macro-parameters such as temperature,
and the environment influences them once it is included in a model.
Similarly, chaotic dynamical systems are highly sensitive to perturbations.
Hence deviations from deterministic trajectories 
are inevitable once a bit of environment exists. 
Dynamics with a finite time singularity \cite{ftsing01}
will inevitably change before the singularity.
Exponentially growing dynamics are likely to meet boundary limitations as well.

Are there deep causal implications of this 
practically unavoidable environmental influence,
particularly when the system happens to be fine-tuned to be influenced?
Does this perceptive condition allow genuine downward causation
from emergent entities? At least, can these queries 
be resolved for emergent phenomena near the corner $Q$, 
where reduction to basic physical causes seems to be assured?
This article starts to address these questions.

Physical reduction is practically ineffectual near the corner $P$.
The renowned biologist Mayr writes \cite[Ch.~4]{Mayr04}:
{\em ``the physiochemical approach is totally sterile in evolutionary biology"},
and {\em "analysis is continued downward only as long as it yields useful new
information"}. Furthermore, Mayr cites Popper's harsh critique \cite[p.~269, 281]{Popper74}: 
{\em ``as a philosophy, reductionism is a failure ... we live in a universe of emergent novelty;
of a novelty which, as a rule, is not completely reducible to any of the preceding stages."}
If physical reductionism is tenable 
after all, it is extremely deeply masked near the corner $P$.
I reckon that this masking is done 
by layers of delegated causality.

\section{Delegated causation}
\label{sec:delegate}

We affirm our focus by the following three definitions.
They 
suggest a non-reductive perspective where the focus 
is on potential interactions between dynamical processes
rather than mathematical behavior of a single dynamical model. 
The first definition characterizes those (continuous or discrete) dynamical systems
or their equilibria, bifurcations, critical phases, ``edge of chaos" conditions 
\cite{Waldrop93}, \cite{Kauffman93} 
that are ``waiting to be perturbed", figuratively speaking. 
\begin{definition} \label{def:primed}
A  dynamical system (or its state) is {\em primed} if it can exhibit complex, 
potentially utilitarian behaviors depending 
on moderate adjustment  of 
boundary conditions.  
\end{definition}
This definition may seem unsatisfactory because of several subjective terms. 
However, the hint of subjectivity and external references are important features of the definition. Primed dynamical systems are to be considered not in isolation
but under influence from each other and the environment. 
They passively, reactively follow the laws of physics, boundary conditions, perturbations.
Most dynamical systems can be considered as {\em primed}
if potential interactions with other systems 
are  sufficiently interesting. Definite examples of non-primed systems would be
chaotic systems or those in a thermodynamic equilibrium. 

Once feedback loops between intricate reactions and triggers 
of various primed systems materialize, an elaborate cybernetic or evolutionary 
system may come out. 
Let us use an economic metaphor for the full spectrum 
of relative, relational configurations of this kind.
\begin{definition}  \label{def:market}
A {\em broad market} is a set of primed dynamical systems,
triggering influences and potential 
reactions between them.
\end{definition}
Extending the economic metaphor,
we may refer to sensitivities of a primed dynamical system
as its {\em demand}, and to the driving effect as {\em supply}. 
A broad market may be structured into hierarchical \cite{Simon62}, \cite{Salthe12}
or cybernetic modules, or dominated by an exchange regime of that supply and demand. 
\begin{definition}  \label{def:cause}
The causal relation between a primed dynamical system and
an external, emergent or self-organized influence that drives the dynamical system 
by a moderate force of interaction is called {\em delegated causation}.
\end{definition}
The ``moderate force" here is defined in the capability context of driving influences.
Their dynamic realization 
is beside the main point.  Just as operative 
features of supply are fairly unimportant in marketing and commerce, 
the causal relation is not defined by the substance of influence factors.
I discuss possible dynamic nature of driving influences 
at the end of \S \ref{sec:yinyang}.

These definitions make most sense for complex systems near the spectral corner $P$
in Figure \ref{fg:2d}. 
Emergent wonders near the corner $Q$ may play special roles in broad markets,
as we can recognize them in computer hardware, inside smartphone screens.
But the meaning of cause delegation within their internal homogeneous ``markets"  
requires further interpretation, as I briefly discuss in the next section.

Delegated causation ought to 
play important roles in realistic complex systems,
and it should be included in the modeling.
Its {\em radical openness} \cite{Chu11} 
discourages modeling of complex phenomena
by single dynamical systems. 
On the other hand, delegated causation allows to explicate cybernetic links,
tipping points, feedback loops. 
The following methodological steps of analyzing  
a single delegated interaction 
must be useful:
\begin{itemize}
\item[\refpart{M1}] 
Identify the primed dynamical system in the interaction.
Determine its sensitivity to perturbations,
and possible reactions to perturbations.
\item[\refpart{M2}] 
Identify the perturbing influences; describe their mechanism.
\item[\refpart{M3}] 
Describe the context, the broad market of the interaction.
\end{itemize}
This methodology is illustrated in Table \ref{tb:m1m3} in \S \ref{sec:biology}
by series of biological examples. The factor \refpart{M2} can be identified
as an {\em interventional cause}. I view \refpart{M1} and \refpart{M2} 
as a dual pair of delegated and interventional causations. 
In the common language, ``delegation" describes plausibly the cases 
when \refpart{M1} is strong, 
while ``intervention" describes the cases when \refpart{M2} dominates. 
I use the language of delegation in the abstract and general sense 
of the causal factor \refpart{M1} in any interaction. 

Layers of  delegated causality signify 
systems with great {\em dynamical depth} \cite{DeaconKou}.
Measures of Kolmogorovian 
{\em sophistication} \cite{sophist} are 
promising for quantifying system complexity. 
Our perspective recommends perturbative analysis 
of dynamical systems and complexity measures,
because significance of primed systems lies in their interactive potential.

Delegated causality amounts to a significant case of Deacon's 
{\em specific absence} \cite{Deacon11}, \cite[p.~119]{Emergence06}
as a ``pulling" causal force. 
Causal roles of constrained absences, virtual demands,
``pregnant" \cite[\S 9]{Salthe12} opportunities merit good appreciation.
For example, early stages of the British industrial revolution were much stimulated
by specific needs of textile, iron, and mining industries \cite{IndustR}.

On the other hand, 
dynamical laws and initial conditions enjoy the respectful status 
of causes. But chaotic dynamics and causality delegation
undermine that status. 
If an attractor, an 
equilibrium, or a self-organized state is reached 
regardless of initial conditions, what is exactly a cause?

\section{Physical reductionism}
\label{sec:reduct}

Openness of delegated causality conforms 
well with Pearl's empirical analysis \cite{Pearl00}
of causation in terms of interventions and counterfactuals.
Interventions are {\em ``actions as external entities, originating from outside our theory, 
not as a mode of behavior within the theory"} \cite{Pearl99}.
Causative interventions on primed dynamical systems 
(and possible malfunctions as reverse interventions)
must be pivotal features of complex systems.

Our first {\em weighty thesis} is this:
delegated causation offers a conceptual mechanism
how micro-scale dynamics results in empirical causation from macro-level
agencies in terms of Pearl's causal calculus \cite{Pearl00} 
and Hoel's {\em causal emergence} \cite{Hoel17}. 
Here is one rationale formulated in the economic terminology:
complex events are better temporally correlated
with appearance of {\em supply} rather than with an onset of {\em demand}, commonly.
A row of falling dominos can be interpreted as a prototypical example.
It would be instructive to relate sensitivities of micro-level dynamics
and {\em effective information} \cite{Hoel17} measures 
in a compelling example of a ``smoothly" emergent phenomenon.

As a mechanism of downward causation, 
delegated causality often entails 
environmental influence. 
Other proposed mechanism of downward causation via environmental interaction 
is {\em practopoietic cycle} \cite[\S 2.6]{Nicolic15}.

Our second {\em weighty thesis} 
refers to G\"odel's incompleteness theorem \cite{Godel31}.
Delegated causality has a self-contradictory flavor 
of a G\"odelian paradox \cite{GodelLiar} in the causality language.
This messes up basic principles 
for physicalism \cite{Kim05}, 
and Kim's argument \cite{Kim98}
against non-reductive physicalism.
In particular,  the {\em causal closure principle} says that 
if a physical event has a cause, it has a physical cause \cite[p.~199]{Emergence06},
and the {\em exclusion principle} states that  
no single event can have two independent sufficient causes \cite[p.~41]{Kim05}.
Kim's conclusion is that non-physical (say, mental) events can have no causal power.
In contrast, delegated causality provokes causal, informational contribution
from external agents or from an emerging organization, even if delegation is not ideal.
Sufficiency of physical causation is debatable then.
Delegated causality clarifies inevitability of causal parity \cite{Weber17} 
and levels of explanation. 
It furnishes hierarchical dynamics, which in turn reinvigorates Aristotle's four categories 
of material, formal, efficient and final causes \cite{Coffman}, \cite[\S 9]{Salthe12}. 

Incidentally, Popper invoked G\"odel's theorem in his argument \cite{Popper74} 
against reductionism. 
By a similar reference to G\"odel's incompleteness, Rosen \cite{Rosen91}
repudiated theoretic formalization of life.
Kim's epiphenomenal implications have been countered by
interventional \cite{ShapiroSober07} 
and counterfactual \cite{ListMenzies09} argumentation.

Causality and reduction in emergent phenomena are customarily analyzed 
in terms of {\em supervenience} \cite{Butterfield12}, \cite{Stalkner96},
\cite[p.~189--243]{Emergence06}:
the principle that entities 
with the same micro-level properties will have the same macro-level properties.
A supervenience is often defined by a coarse-graining map 
from physical micro-states to emergent macro-level states.

Delegated causality amounts to engaging external information, 
possibly entailing an {\em externalist} \cite{External}, \cite{Menary10}
propensity to expand the supervenience base of micro-states
beyond memorization, representation. 
This propensity can be quickly consummated in complex systems near the corner $P$,
making a supervenience analysis doubtful. 
For example, an animal may habitually follow 
certain external clues, peer group behavior or ``expert" guidance.
Or  the neuro-physiological basis of its behavior 
might be ``eagerly" changing in live action.
These are examples of {\em situated}, {\em embodied} cognition \cite{MWilson02}.

Human consciousness and free will are the most prominent emergent phenomena. 
Without more ado, they could be viewed as pinnacles of delegated causation
in the known cosmos. 
From a utilitarian perspective, consciousness is a cognitive-behavioral characteristic 
that is able to intervene on (sometimes particularly quiet) emotional, somatic drivers.

Can emergent systems near the corner $Q$ in Figure \ref{fg:2d}
be interpreted as ``eagerly" seeking outside influence, 
even if the outside is presumably negligible?
In a sense, phase transitions delegate causality to dust particles, matter irregularities.
Most dramatically, we can consider the whole Universe with no outside in principle.
Can we then speculate that a collective behavior of the system is 
seeking to {\em externalize} its statistical parameters, 
thermodynamic ``forces" \cite{Onsager31},
thereby meeting Mach's principle \cite{Mach}, 
renormalization group dynamics \cite{Batterman}, \cite[Ch.~3]{Mainwood06}, 
Heisenberg's indeterminacy, or an observer at the ``boundary"?
Can dynamic novelty be a valid extension of a supervenience basis?
Does emergence itself reflect primordial ways 
of conceding causality? 
At least, these speculations could be simulated by a chaotic, fractal 
or hardly computable topology of the supervenience map,
while the macro-dynamics would be described by smooth functions. 
The consideration of emergence as delegated causality is 
supported by the recently established correspondence
between the renormalization group in theoretical physics
and the deep learning approach in artificial intelligence
\cite{Wolchover14}, \cite{MehtaSchwab};
I elaborate this in \S \ref{sec:integrate}.

The above questions ought to be addressed by 
{\em theories of everything} \cite{Hawking05}.
Delegated causality gives an apparent 
taste of physical implications of %
G\"odel's incompleteness theorem 
\cite{Jaki66}, \cite{DysonReview}, \cite{HawkingTOE}.
It 
may even turn out to be a reformulation
of G\"odel's incompleteness in causal terms.

The notion of delegated causality enhances 
rather than renounces the reductionist paradigm
by defining 
the causal role of critical, ``edge of chaos" systems.
Possibilities of extravagant dynamics are wholly controlled by micro-level arrangements.
But dynamical actualization is contingent to 
particular instabilities or input from the environment.

\section{Biological causality}  
\label{sec:biology}

Biology is a great ground for testing explanatory power of delegated causality. 
We can expect lots of elaborate causative interventions, provocations.

Mayr's influential article \cite{Mayr61}
distinguishes {\em proximate} (mainly physiological)
and {\em ultimate} (mainly 
evolutionary) causes of biological phenomena. 
Contrary to the reductionist template, 
the relatively more ``teleological" level 
of explanation by natural selection and adaptation 
is considered more fundamental in the conventional 
(neo-Darwinian) Modern Synthesis. 
Physiological development is guided 
by the genetic code, which in turn is pressured by natural selection.
By 
a rigid interpretation of Modern Synthesis, 
biological functionality and organisms can be fully understood 
only from the evolutionary perspective, 
while comprehension of ontogenic development 
is principally unnecessary for that.
This {\em reduction} of biological causes 
to statistical phenotype selection, genetic adaptation and drift 
could be a deep reason of steady critique of the Modern Synthesis 
\cite{Nature14}, \cite{Welch17}, \cite{Gould80}, \cite{PigliucciMuller},  
\cite{VSD13}. 

Laland and co-authors 
\cite{Laland11e}, \cite{Laland13e} suggest 
that Mayr's proximate-ultimate dichotomy, 
although still vital, hinders a proper integration of evolution and development,
recognition of multiple sources of evolutionary novelty. 
They advocate an intimate relation between developmental and evolutionary processes,
with the former able to influence evolutionary change through phenotypic plasticity, 
developmental bias, epigenetic inheritance, behavioral changes, 
and ecological interactions such as niche construction. 
This is a good list to test scientific productivity of the delegated causality notion.
Genes are masters of causal intervention in the biological world,
enjoying vast biochemical infrastructure.
But they themselves may be 
open to interventions. Feedback from 
developmental and ecological conditions would be 
a powerful source of adaptation, with diversified 
agencies 
and information forms as inevitable effects.

\begin{table} \small
\begin{tabular}{lll}  \hline
\refpart{M1} & \refpart{M2} & \refpart{M3} \\
 \hline \\[-10pt] 
cell & cardiovascular system & supply of oxygen, nutrients \\[1pt] 
cardiovascular system & heart & blood circulation \\[1pt] 
tissues, organs & endocrine system & hormonal coordination \\[1pt] 
pathogenic infection & immune system & protection from diseases \\[1pt] 
\hline \\[-10pt] 
host organism & parasite & parasitism \\[1pt]
prey & predator & food chain \\[1pt]
ecosystem & invasive species & ecological disruption \\[1pt]
habitat & niche construction & adaptive alteration  \\[-1pt]
& & \qquad of the environment \\[1pt]
\hline \\[-10pt] 
species population & mountainous topography & geographic 
speciation \\[1pt]
heritable variation & differential reproduction & natural selection \\[1pt] 
gene pool & sexual reproduction & increased genetic variation \\[1pt]
genetic drift & punctuated equilibrium & macro-evolution \\[1pt]
\hline \\[-10pt] 
chemical compounds & enzyme & catalyzed reaction \\[1pt] 
nucleic acids & ribosome & protein synthesis \\[1pt]  
allosteric enzymes & inhibitory or activating &
regulation of metabolic \\[-1pt] 
& \qquad metabolites 
& \qquad pathways \\[1pt]  
reactants with & 
non-covalent  bonding  & versatile stereospecific   \\[-1pt]
\quad complementary sites & 
& \qquad discrimination \\[1pt]  
oligomeric proteins & stereospecific bonding & spontaneous self-assembly \\[1pt]
stereospecificity & chemical potential & information amplification \\[1pt] 
\hline \\[-10pt] 
fertilized egg & genome 
& ontogenic development \\[1pt] 
genome & gene 
regulatory network & morphogenesis \\[1pt]
stem cells & variable gene expression & cell differentiation \\[1pt] 
germ layers & cell sorting, EMT & tissue separation \\[1pt]
\hline
\end{tabular}
\caption{Biological examples of 
delegated causality  analysis, as outlined 
 in \S \ref{sec:delegate}.}
\label{tb:m1m3}
\end{table}

The methodology \refpart{M1}--\refpart{M3} in \S \ref{sec:delegate} 
of explicating delegated causation is helpful to 
highlight interventional forces \refpart{M2}
and signification \refpart{M3} of interactions. 
This is illustrated in Table \ref{tb:m1m3} 
by several physiological, ecological, evolutionary, biochemical \cite{Monod72}, 
and developmental \cite{Fagotto14}, \cite{emt16} examples.
In many cases, the dual force of intervention dominates. 
But the central role of genes is compellingly 
a cybernetic hub of delegation, as I recount again in the middle of \S \ref{sec:integrate}.
In the biochemical context, the columns \refpart{M1}, \refpart{M2} 
delineate Rosen's rendition \cite{Rosen91}
of Aristotle's material and efficient causes, respectively.
Systemic closure of efficient causes is underscored by 
theories of biological autonomy \cite{MorenoMossio}, \cite{Autopoiesis}.
In the context of causal stability and specificity \cite{Calcott17}, \cite{Woodward10}, 
the factor \refpart{M1} tends to provide specificity,
while \refpart{M2} furnishes stability and permissiveness.
The analysis \refpart{M1}--\refpart{M3}
must offer more clarity than the proposal in \cite{Laland11e}, \cite{Laland13e} 
to employ a terminology of {\em reciprocal causation}.
The polarity between Mayr's proximate and ultimate causes 
should rather stay.

Other biological disciplines where explicit analysis of delegated causation 
ought to be useful
are microbiology, symbiosis \cite{symbiosis}, 
communication \cite{biocomm}. 
The contentious subjects of group selection \cite{WilsonSober94}, 
multilevel selection \cite{Okasha06}, evolvability \cite{EvoDy03}, 
cooperation \cite{Nowak06}, altruism \cite{SoberWilson88}, 
longer term adaptation 
might be greatly altered by the perspective of delegated causation as well. 
The criticized theory of Wynne-Edwards \cite{WEdwards62}
of territorial and hierarchical 
organizations in various species regulating population growth 
becomes more plausible, 
as game-theoretic explanations \cite[Ch.~7]{DawkinsSG}  turn 
into proximate causes relative to the ``ultimate" 
adequacy between population size and resources.
Mayr \cite[Ch.~8]{Mayr04} acknowledges selection of cohesive social groups
because their fitness values can be disproportionately larger 
than the mean values of individual fitness.
This is reminiscent to the emergence principle that 
{\em the whole is greater than the sum of its parts}.

\section{Yin and Yang}
\label{sec:yinyang}

Conceptions resonating with {delegated causality} can be found in Chinese philosophy.
Provocation of some experiential apprehension from readers looks 
sensible 
or frankly unavoidable in this presentation. 
That being the case, this section could be excused for being a little rhetorical. 
Certain cultural, social parallels may be promptly triggered.
Carefully evaluative readers are encouraged to 
contain social, emotional charges and judgements.


The ancient Chinese concepts 
 of Yin and Yang are customarily 
evoked to affirm complementarity of opposing, interdependent forces. 
It is less widely known that Yin and Yang are defined \cite{OxED}
as two concrete complementary principles.
The complementary harmony is better underscored by Taoism \cite{CapraTao},
while the particular duality is more emphasized by Confucianism \cite{Rosenlee07}. 
Recently, the Yin-Yang polarity has been interestingly related to
holistic causality \cite{Chung16}, 
epistemology \cite{Zhang07}, \cite{Benetatou16},
transformational change \cite{KleinWong12}, \cite{Li2008}, and even 
microbiology \cite{Zhang14}, molecular biology \cite{YY1}.

One of the defining complementarities is that Yin is an 
abstraction  of {\em passivity, inertness}, 
while Yang is the {\em active, generative} principle. 
This particular duality captures the contrast between mechanical dynamical systems
and emergent phenomena very well. 
A primed dynamical system of Definition \ref{def:primed}
would be a passively reactive Yin,
while any effective influence 
in Definition \ref{def:cause} would be Yang.
It seems fittingly convenient to adopt 
this language.


Paraphrasing an earlier 
statement, dynamic realization of Yang is beside the main point.
Yang is the novelty, emergence beyond Yin's territory. 
Relative to underlying Yin dynamics, 
Yang is a hero \cite{Campbell}, a master, an artful trickster.
Yang is the magic of actualizing synergetic possibilities.
Yang ``opposes" entropy increase by causative leadership,
blocking unwelcome occurrences, 
forcing a decisive turn of action,
claiming the language of communication. 

It is not necessary 
to assume supernatural causes to explain Yang manifestations.
Yin's ready anticipation of external perturbation and
yielding to self-precipitated pressures
are {\em pulling} causative forces already.
Yang is defined by this anticipatory perception of Yin. 
We may refer to a finely triggered Yin's incipient reaction
as {\em Yin's satisfaction}. 
In some cases, this satisfaction may be prompted straightforwardly; 
in other cases it may be an uncertain, rare 
event. Both Yin and Yang of a particular interaction may evolve
to sophisticated cybernetic levels, 
meeting 
criteria of a broad market.

With many Yin agents available as amenable resources, 
Yang competition to employ them emerges. 
Or who harnesses whom? 
Attribution of selection powers is relative in multifaceted 
interactions of Yin and Yang. 
There is a certain mutual inclusion of Yin and Yang,
symbolized by the taijitu sign: 
\[ 
\includegraphics[width=100pt]{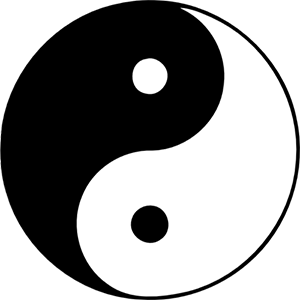} \]
For 
example, a perceptual mechanism is Yin
primed to be affected by the environment.
But Yang should have perceptual capabilities
to recognize peer Yin's dynamic demand and address it.

The simplest mechanism of Yang (as external dynamics) is randomness.
This means, Yin allows chance to direct its dynamics.
For example, divination rituals were common in ancient human societies.
In China of the Shang dynasty (circa 1600 BC$\,-\,$1046 BC),  
oracle bones of ox or turtle were prominently used \cite{OBones}.
Other simple mechanism for Yang is competition.
That is, Yin sets up a contest who best performs or fakes a particular demand.
Another manifestation of Yang is information.
Generally, Yang can be identified in processes
that are called ``teleological" now and then. 
It is instructive to inspect 
Mayr's classification of five teleologies \cite[Ch.~3]{Mayr04}:
{\em ``four of the five phenomena traditionally called teleological
can be completely explained by science, while the fifth one, 
cosmic teleology, does not exist."}
The similar list of Ellis \cite{Ellis12}, \cite[Ch.~4]{Ellis16}
of five downward causations is definitely germane as well.

\section{An integrated view}
\label{sec:integrate}

The two {\em weighty theses} in 
\S \ref{sec:reduct} relate delegated causality 
to emergence and G\"odel's incompleteness theorem.
Emergence and G\"odel's  
theorem are brought together 
in the literature at times \cite{Requart91}. 
For example, Jorgensen and Svirezhev write \cite[p.~8]{Thermoeco04}:
{\em ``In accordance with G\"odel's Theorem, the properties of order and emergence
cannot be observed and acknowledged from within the system, but only by an outside observer."} It could be meaningful to consider 
emergent phenomena and 
selection modes as physical, biological, or socionomic \cite{Prechter03} 
manifestations of G\"odel's incompleteness.
Monod's concept of {\em gratuity}  \cite[Ch.~IV]{Monod72} ---
i.e., independence between chemical qualities and function
of biochemical processes --- can be similarly related to G\"odel's theorem as well. 

Delegated causality facilitates hierarchical dynamics
and defines {\em Nature's ``joints"} \cite[\S 4]{Salthe12}
along which dynamical levels are fused.
Higher levels impose boundary conditions, 
constrains, selection regimes on dynamics of lower levels \cite{Juarrero99}. 
This downward causation is enabled by critical, primed organization
of the lower levels. 
Intermediate ``sub-wholes" in a dynamical hierarchy
are Koestler's {\em holons} \cite{Koestler78}
--- i.e., stable, integrated, largely autonomous, yet interactive entities. 
They are both primed dynamical systems and interventional forces, 
both consumers of 
energy and local sources of order.
The constrains of a dynamical level  define distinct dynamics, 
bias, functionality, and a ``behavior code" of the holons.

Recently \cite{MehtaSchwab}, \cite{Wolchover14}, 
an equivalence between the renormalization technique in condensed matter 
physics and the deep learning approach in artificial intelligence was established.
This reinforces the interpretation that micro-dynamics of a phase transition
(or a dissipative system such as B\'enard's cells \cite{Benard81}, \cite[p.~22--25]{Swenson97})
is organizing itself to ``explore" and adjust to macro-boundaries. 
The scale-free dynamics extends mean free-path distances 
and relaxation times of particle interactions by orders of magnitude, 
until limiting macro-dimensions are eventually met.
Causality delegation becomes a paraphrase of this ``deep learning" of macro-dimensions.

This enriched metaphorical context echoes 
the Santiago school view of living systems as cognitive \cite{Autopoiesis},
sense making \cite{Thompson} systems,
and Heylighen's 
view \cite{Hey11}  of evolution and self-organization as cognitive processes. 
The action-centered ontology \cite{Hey11} assigns 
{\em intentional stance} \cite{Dennett87} to 
increasingly adaptive agents, and (in essence) recognizes
interventions as basic constituents of reality.
But the passive, incomplete kind of delegated causality should be recognized as well.
Deacon's \cite{Deacon11} cryptic notion  of {\em ententionality}  is relatable here,
if it characterizes being organized for some functioning.
Elements of {\em anticipation} relate primed dynamical systems, deep learning,
and {\em enactive} \cite[Ch.~8]{VarelaTR} cognition.

Primed dynamical systems and delegated causality ought to be discerned 
in all phenomena and entities that are considered emergent or self-organizing.
For example, a living cell is predominantly a product of developmental processes
that delegate their coordination to genes. 
Biochemical reactions in the cytoplasm are orchestrated by genes
in the form of nucleic acids, 
but epigenetic switching, energy input from mitochondria, and nutrient flow  
determine the mode of metabolism. 
Arrangements of tissues and organs are coordinated by genes again,
but the behavior and fate of organic ``vehicles" \cite[\S 12]{DawkinsSG} 
are delegated to the nervous system, sociality, tribal customs, 
or eventually to democratic politics.
 Natural, symbiotic, artificial, and cultural selections blend 
 into a rather continuous spectrum of efficient causes.
  
Intuition on delegated causality can be further enhanced by reflecting on
many regarded modes of economic causality 
\cite{Hoover01}, \cite{MorckYeung11}, \cite{Varian16}, \cite{Joffe17}.
Furthermore, 
it is worth to reflect on causality of avalanches 
in self-organized critical systems \cite{Bak96}, \cite{SOC15}, 
where (in theory) only a scale-free statistical distribution has a predictive power.
Scale-free dynamics in heterogeneous systems 
near the corner $P$ in Figure \ref{fg:2d} can be self-reinforced
by a few persistent motives of adaptation throughout the scale expanse.

The Chinese concepts of Yin and Yang describe dynamical discontinuities fittingly. 
The interaction between Yang's constraints and Yin's demand or satisfaction should define
a statistical partition for a relevant entropy tally, and 
the semantics of what is caused by delegation. 
Yang's contingency suggests that universal principles 
of self-organization or non-equilibrium thermodynamics 
(such as speculated laws of {\em maximal entropy production} \cite{Swenson97})
would rather describe potentialities.

Overwhelming interventions, catastrophes, 
dynamical collapses, {\em black swans} \cite{Taleb07} 
fit the presented context of delegated causality as Yang forces.
Resignation to them would not be called ``delegation" in the common language.
On the other hand, ignoring predictably unsustainable 
trends (even if the timing and operation of a likely resolution is highly uncertain) 
amounts to delegation of responsibilities for consequences.
For example, human conscious effort to address a climate change \cite{Zhang11}
is a potentially significant causal factor. Likelihood of a catastrophe
indicates 
a primed dynamical system, to be influenced possibly by its constituents.


The perspective of Yin and Yang applies 
to the cultural discord between conservative and post-modernist \cite{foucault80}
views in the United States. 
As the American academic critic Bloom noted \cite[p.~25]{Bloom87},
the post-modernist 
{\em ``relativity of truth is not a theoretical insight but a moral postulate"}.
Post-modernism can be viewed 
as celebration of Yin birthrights and subjective preferences, as well as 
denial of Yang authority.
Firstly however, limitation of resources is not addressed operatively
by the progressive optimism. That is left implicitly to social, political, 
or financial \cite{Graeber11} hierarchies, be them patriarchal or not. 
Congruent resolution of oppression and sustaining social progress 
may continue to be 
historical challenges  \cite[p.~183]{Graeber11},
especially under a prolonged environmental stress.
Secondly, Yang standards of leadership are in actuality yet appreciated
by the progressive 
electorate. This is evident from the 2016 US presidential election \cite{Trump16},
where Hillary Clinton did not fully 
motivate her expected voters.

\section{In conclusion}

Common experience, biology \cite{Mayr04}, empirical causality \cite{Pearl00}, 
information measures \cite{Hoel17}, \cite{TononiSporns03}
tell that downward causation apparently exist.
The interpretation of the ancient 
Chinese concept of Yin as a primed dynamical system 
intimates 
that downward causation does exist in a very strong, virtually mechanical sense.
Instead of having 
{\em ``higher scales wrest the controls from lower scales"} \cite{Wolchover17},
the {\em lower scales} can be organized to operatively concede 
a good deal of causality to some {\em higher scales}.
This possibility shifts the reductionistic paradigm in a novel way \cite{Kuhn70}. 
The duality of delegated and interventional causations is 
subtly eminent in Yin-Yang philosophy. 
Letting things go their way for a matched 
balance is a part of this duality.

Co-evolution of agencies with expanding 
capabilities  to provoke, perceive, impose causal relations
can lead to understandably tremendous consequences.
Conspicuously complex natural systems build up on 
interactions of delegated causality.
This must be it. 

\section*{Acknowledgements}

This article is influenced by informed experience
of personal development since 2012. 
The understanding of Yin and Yang is inspired 
by works of David Deida, Adam Gilad, David Shade. 
The emphasis on causality sank in while listening to Jason Capital, Bobby Rio. 
For an introduction to externalist, interventional,
mythological perspectives, thanks to Joseph Riggio.
Constructive comments from Ari Belenkiy, Marianne Benetatou, James Read, 
Stanley Salthe,  Susumu Tanabe, Alexander Tokmakov, 
and fruitful conversations with Yang-Hui He, Minxin Huang,
Sanjaye Ramgoolam, George Shabat
are much appreciated.

\renewcommand{\baselinestretch}{0.98}

\end{document}